\documentclass[fleqn,usenatbib,useAMS]{mnras}

\usepackage{graphicx}	
\usepackage{amsmath}	
\usepackage{amssymb}	
\usepackage{multicol}        
\usepackage{bm}		
\usepackage{pdflscape}	

\usepackage[T1]{fontenc}
\usepackage{ae,aecompl}

\newcommand{\bc}{\begin{center}}
\newcommand{\ec}{\end{center}}

\newcommand{\hMsun}{~h^{-1}\>{\rm M_\odot}}

\newcommand{\Hunit}{~h ~{\rm km}~s^{-1}~{\rm Mpc}^{-1}}

\defcitealias{Cui2015}{Paper~I}

\title[On The Theoretical Cluster Dynamical State]{
  On the dynamical state of galaxy clusters: insights from
  cosmological simulations II.}

\author[Weiguang Cui et al.]
{\parbox{\textwidth}{Weiguang Cui,$^{1,2}$\thanks{E-mail: \texttt{weiguang.cui@uwa.edu.au}}
  Chris Power,$^{1,2}$ Stefano Borgani,$^{3,4,5}$
  Alexander Knebe,$^{6,7}$ Geraint F. Lewis,$^8$ Giuseppe Murante,$^3$ and
  Gregory B. Poole,$^{9}$ }\vspace{0.4cm}
  \\
\parbox{\textwidth}{
  $^1$ ICRAR, University of Western Australia, 35 Stirling Highway,
  Crawley, Western Australia 6009, Australia\\
  $^2$ ARC Centre of Excellence for All-Sky Astrophysics (CAASTRO)\\
  $^3$ Astronomy Unit, Department of Physics, University of Trieste,
  via Tiepolo 11, I-34131 Trieste, Italy\\
  $^4$ INAF -- Astronomical Observatory of Trieste, via Tiepolo 11,
  I-34131 Trieste, Italy\\
  $^5$ INFN -- Sezione di Trieste, I-34100 Trieste, Italy\\
  $^6$Departamento de F\'isica Te\'{o}rica, M\'{o}dulo 15, Facultad de Ciencias,
  Universidad Aut\'{o}noma de Madrid, 28049 Madrid, Spain\\
  $^7$Astro-UAM, UAM, Unidad Asociada CSIC\\
  $^8$ Sydney Institute for Astronomy, School of Physics, A28,
  The University of Sydney NSW 2006, Australia \\
  $^{9}$ School of Physics, University of Melbourne, Parksville,
  VIC 3010, Australia
}}

\date{Accepted XXX. Received YYY; in original form ZZZ}

\pubyear{2015}

\begin{document}
\label{firstpage}
\pagerange{\pageref{firstpage}--\pageref{lastpage}}
\maketitle

\begin{abstract}
  Using a suite of cosmology simulations of a sample of $\textgreater 120$ galaxy clusters with
  $\log(M_{DM, vir}) \le 14.5$. We compare clusters that form in purely dark matter run
  and their counterparts in hydro runs and investigate
  4 independent parameters, that are normally used to
  classify dynamical state. We find that the virial ratio $\eta$ in
  hydro-dynamical runs is $\sim 10$ per cent lower than in the DM run, and
  there is no clear separation between the relaxed and unrelaxed clusters for
  any parameter. Further, using the velocity dispersion deviation parameter $\zeta$,
  which is defined as the ratio between cluster velocity dispersion $\sigma$ and
  the theoretical prediction $\sigma_t = \sqrt{G M_{total}/R}$, we find that
  there is a linear correlation between the virial ratio $\eta$ and this $\zeta$
  parameter. We propose to use this $\zeta$ parameter, which can be easily derived
  from observed galaxy clusters, as a substitute
  of the $\eta$ parameter to quantify the cluster dynamical state.
\end{abstract}



\section{Introduction}
\label{i}
Currently favored models of cosmological structure formation are hierarchical --
lower mass systems merge progressively to form more massive structures, with
galaxy clusters representing the final state of this process. The dynamical
process, driven by gravity, determines the final properties of the
dark matter halo, as well as the baryonic contents in it -- galaxies, intra
cluster medium (ICM), etc. However, even at the final state of hierarchical
structure formation, the galaxy clusters are not always in dynamic equilibrium. In
observations, galaxy cluster systems can be roughly separated
into relaxed and unrelaxed; the ICM in relaxed clusters is normally in
hydrostatic equilibrium, while dynamically unrelaxed clusters are
undergoing, or have undergone, a merger, which leaves
the ICM turbulent \citep[see][and references therein]{Wen2013}. In simulations,
there are a vague of ways dynamical state can be evaluated.

Using dark-matter-only simulations, \cite{Jing2000} found that about $30$ per cent of
the simulated dark matter halos can not be fitted by the NFW profile \citep{NFW},
and these halos that showed larger deviations from the NFW profile exhibited
significant internal substructures.
Using the integral virial ratio parameter $2T/|W|+1$, here T is the kinetic
energy, W is the potential energy, \cite{Bett2007} suggested $2T/W +1 \textless 1.5$
to select halos in quasi-equilibrium states \citep[see also][]{Klypin2016}.
\cite{Neto2007} expanded the criteria by including substructure mass fraction
and centre-of-mass offset. However, they adopted a narrower limit for their virial
ratio $2T/|W| < 1.35$ \citep[see also][]{Ludlow2012}. \cite{Shaw2006,Poole2006,Davis2011}
modified the virial ratio by taking the surface
pressure energy $E_s$ into account. This is because halos are not isolated
in cosmology simulations, and infalling materials alter $2T/W$. Besides the surface
pressure energy, \cite{Davis2011} also considered the potential energy from particles
outside of halos -- $W_{ext}$ for the virial ratio. However, they
found that $W_{ext}$ is negligible. Nevertheless,
different limits are used to calculate the virial ratio:
$(2T-E_s)/W +1 > -0.2$ for \cite{Shaw2006}; $|1+2T/(E_s + W)| < 0.02$
for \cite{Poole2006}; While \cite{Knebe2008} suggested $-0.15 \leq (2T-E_s)/W +1
\leq 0.15$ (with a mass dependence at $z$ = 1) to select out relaxed halos.
\cite{Power2012} studied the relation between centre-of-mass offset
and equilibrium state. Instead of using virial ratio, they suggested a centre
of mass offset value of $0.04$ to select relaxed halos.

All of these studies were based on dark matter only simulations. However, as
numerical simulations with sophisticated sub-grid baryon models have become
more mature and successful in producing observed-like galaxies, there has been
great interest in studying the baryonic effects on galaxy cluster properties
\citep[e.g.][]{Schaller2015b,Cui2016}; on power spectrum
\citep[e.g.][]{Daalen2011}; on halo mass as well as halo mass function
\citep[e.g.][]{Cui2012a,Cui2014b,Velliscig2014}; and on substructure shapes and
alignments \citep[e.g.][]{Knebe2010,Velliscig2015}. It is timely and interesting
to study and how baryons affect the dynamical state of galaxy clusters. Baryons,
especially gas, are subject to other forces in addition to gravity to dark
matter, which will lead changes on T and W.

In this paper, we study the dynamical state of galaxy clusters with a volume- and
mass-complete sample from a series of cosmological simulations
with three different baryon models, which we have presented in
\citet[][hereafter Paper~I]{Cui2015}. We investigate how different measures of dynamical state change between dark-matter-only and hydro-dynamical runs.

In the following sections, we briefly describe these hydro-simulations with
different baryon models \citep[see also ][]{Cui2012a, Cui2014b} and the
statistical sample of clusters \citepalias[see also][]{Cui2015}
(\S~\ref {simulation}), and present our dynamical state classification methods
(\S~\ref {method}). In section~\ref {results} we present our results.
Finally, we summarise our conclusions in \S~\ref {concl}, and comment on the
implications for interpretation of observations of galaxy clusters.

\section{Simulated Galaxy Cluster Catalogue}
\label{simulation}

These simulations use a flat $\Lambda$CDM cosmology, with cosmological
parameters of $\Omega_{\rm m} = 0.24$ for the matter density parameter,
$\Omega_{\rm b} = 0.0413$ for the baryon contribution, $\sigma_8=0.8$
for the power spectrum normalization, $n_{\rm s} = 0.96$ for the primordial
spectral index, and $h =0.73$ for the Hubble parameter in units of $100 \Hunit$.
They used the same realization of the initial matter power spectrum,
and were run with the TreePM-SPH code {\small GADGET-3}, an improved version of the
public {\small GADGET-2} code \citep{Gadget2}. Three simulations were run, we
refer to the dark-matter-only simulation as the DM run;
the hydrodynamical simulations including radiative cooling, star formation and
kinetic feedback from supernovae: in one case we ignore feedback from AGN (which is
referred as the CSF run), while in the other we include it (which is referred as
the AGN run). The DM run has two families of dark matter particles: the one with
larger particle mass shares the same ID as the dark matter particles in the CSF
and AGN runs; while the one with smaller particle mass has equal mass as the gas
particles in the CSF and AGN runs at the initial condition of $z$ = 49. With this
particular setup, we can make an explicated investigation on the baryon effect.

Halos are identified using the Python
spherIcAl Overdensity (SO) algorithm {\textsc PIAO}
\footnote{It is publicly available at https://github.com/ilaudy/PIAO}
\citep{Cui2014b}, and are selected from the DM run with a mass cut.
We reselect 123 halos, which have the virial mass of $\log_{10} (M_{vir}) > 14.5
\hMsun$. We use \cite{Bryan1998} t oestimate $\Delta_{vir}$ and compute $M_{vir}$.
Counter parts SO halos in AGN and CSF runs are identified by cross-matching
dark matter components using their unique particle IDs
\citep[also see][more details]{Cui2014b}.

\section{Methods}
\label{method}


\paragraph*{Virial Ratio}
The exact virial theorem for a self-gravitating system is
  \begin{equation}
    \frac{1}{2} \frac{d^2 I}{dt^2} = 2T + W - E_s,
    \label{eq:vt}
  \end{equation}
where I is the moment of inertia. The proper way of calculating the equation
\ref{eq:vt} is by using the time averaged values of these quantities
\citep[see discussion in][]{Poole2006}. However, due to the limited
outputs of the simulation, we only calculate these quantities at $z$ = 0.

Total kinetic energy T is calculated differently
for collisionless (dark matter and star) particles and collisional (gas) particles.
After removing the halo motion, which is given by the mass-weighted mean velocity from
particles within $30 {\rm kpc}$, and the Hubble flow, T is simply $\frac{1}{2}m_i v_i^2$,
where $i$ is for all collisionless particles; We use the gas thermal energy U for its
kinetic energy. Total potential energy W is directly calculated by using all particles
inside halos without any approximation. $E_s$ is the energy from surface pressure P at the
halo boundary. As described in \cite{Chandrasekhar1961}, $E_s$ is
  \begin{equation}
    E_s = \int P(r){\mathbfit{r}} \cdot {\mathbfit{dS}}.
  \end{equation}
Assuming the ideal gas law, P for collisionless particles \citep[see][for more details]{Shaw2006}
can be written as
  \begin{equation}
    P_c = \frac{\sum_i m_i v^2_i}{3 V},
  \end{equation}
this summation is over all particles with mass $m_i$, velocity $v_i$ inside volume V,
while P for gas particles \citep[see][for more details]{Poole2006} is
 \begin{equation}
    P_g = \frac{\sum_i N_i k_B T_i}{V},
  \end{equation}
here $N_i$, $T_i$ are the gas number and temperature respectively, $k_B$ is the Boltzmann
constant.

We follow \citet{Shaw2006} to calculate P: first, we rank order all
particles by their radius and select the outermost 20 per cent; then we
label the radius of the innermost particle in this shell as $R_{0.8}$, the
outermost as $R_{vir}$, and the median as $R_{0.9}$. V is the volume
occupied by the outermost 20 per cent particles,
$V = \frac{4 \upi}{3} (R^3_{vir} - R^3_{0.8}).$
The surface pressure energy from collisionless component can be approximated by
  \begin{equation}
    E_{s,c} \approx 4 \upi R^3_{0.9} P_c = \frac{R^3_{0.9}}{R^3_{vir} -
    R^3_{0.8}} \sum_i m_i v^2_i.
  \end{equation}

For gaseous particles, the gas number density n can be expressed in terms
of the mean molecular weight: $\mu = {\rho}/(n m_p)$,
where $m_p$ is the mass of a proton, $\rho$ is gas density. Following \citet{Mo2010},
we assume the elements heavier than helium have a mass number $M_i \approx 2(Q_i +1)$,
here $Q_i +1$ is the charge number of a fully ionized atom. If we
define the total mass as $X_i = 1$, where $X_i$ is the mass abundance of
element i, then we have $\mu = {4}/(6 X_H + X_{He} + 2)$.
Normally we assume the metallicity $Z = 1 - X_H - X_{He}$ is very small, and
the mass fraction for hydrogen is around 0.76. Thus, we can have
$\mu \approx 0.588$ and the gas number
\begin{equation}
N = nV = \frac{\rho V}{\mu m_p} = \frac{m}{\mu m_p}.
\end{equation}
Finally, we can calculate the surface pressure energy from the gas component as,
\begin{equation}
  E_{s, g} \approx 4 \upi R^3_{0.9} P_g = \frac{R^3_{0.9}}{R^3_{vir} - R^3_{0.8}} \frac{3
    k_B}{\mu m_p} \sum_i m_i T_i,
\end{equation}
where summation is over all the gas particles lying between $R_{0.8}$ and $R_{vir}$.
$E_s$ is contributed by both collisionless and gas particles.

If the system is in a steady state and dynamical equilibrium,
equation~(\ref{eq:vt}) will reduce to $2T + W - E_s = 0$, which can be rewritten as
$(2T - E_s)/|W| = 1$. Therefore, we define $\eta = (2T - E_s)/|W|$,
and expect $\eta \rightarrow 1$ for dynamically relaxed galaxy clusters.

\paragraph*{Total subhalo mass fraction} Subhalos are identified by
\textsc{SubFind} \citep{Springel2001, Dolag2009, Cui2014a}.
For all the galaxy clusters identified by \textsc{PIAO}, we run
\textsc{SubFind} on them one by one. The smallest subhalo has at least
32 particles. Subhalos with only gas particles are not taken into account
\citep{Dolag2009}. The subhalo mass fraction $f_s$ does not include the most
massive substructure as this is simply the bound component of the main halo.

\paragraph*{Centre-of-mass offset} For all the particles within the virial radius
$R_{vir}$, we compute the centre of mass as
\begin{equation}
  {\mathbfit{R}}_{cm} = \frac{1}{M} \sum_{i=1}^n m_i r_i,
\end{equation}
here $m_i$ is the $i^{th}$ particle mass, $r_i$ is its position, M is the halo
virial mass, and n is the total particle number within $R_{vir}$.
The centre offset is defined as
$\Delta_r = |{\mathbfit{R}}_{cm} - {\mathbfit{R}}_c|/{R_{vir}}$.
We note here that the density peak position is used as cluster centre
${\mathbfit{R}}_{c}$ \citepalias[see][for more discussion about different
centre definitions]{Cui2015}.

\paragraph*{Velocity dispersion deviation}
The velocity dispersion $\sigma$ is always an important quantity for cluster
dynamics. It is often used to predict the cluster's dynamical mass through the
virial theorem:
\begin{equation}
  \frac{1}{2} M_{total} \sigma^2 \propto \frac{G M_{total}^2}{R},
\end{equation}
where G is the universal gravitational constant, and $M_{total}$ and R are the cluster
mass and radius. Thus, one can easily get the predicted dynamical mass through
$M_{total} \propto (R \sigma^2)/{G}$. However, this is based on the assumption
that the cluster is in dynamical equilibrium, which is normally not true. Therefore,
we define a parameter to quantify the deviation to the dynamical equilibrium:
$\zeta = \frac{\sigma}{\sigma_t}$, here $\sigma_t = {\sqrt{(G M_{total})/{R}}}$.
Note that the velocity dispersion deviation $\zeta$ can be
different from unity even for perfectly relaxed clusters,
because its exact value also depends on the density profile.

\section{Results}
\label{results}

\subsection{Radial profiles}
\label{4.1}
\begin{figure*}
\includegraphics[width=\textwidth]{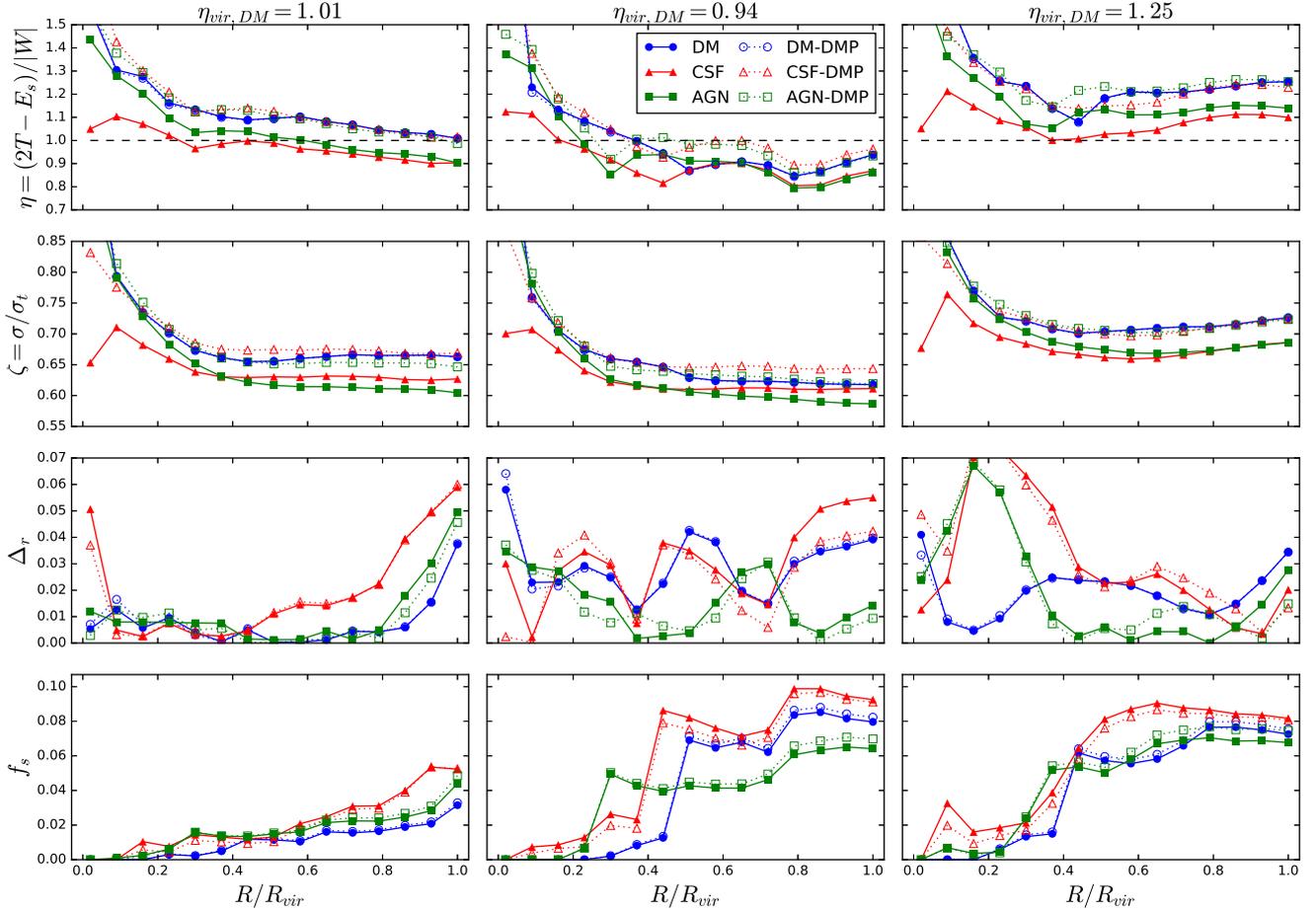}
\caption{
  The values of $\eta$, $\zeta$, $\Delta_r$, $f_s$ (from top to bottom panels) as
  a function of normalized radius out to $R_{vir}$. From left to
  right columns, we show the three example clusters with different $\eta_{DM}$
  values at $R_{vir}$: $\eta_{vir, DM} = 1.01, 0.94, 1.25$, respectively.
  As shown in the legend on the top middle panel, blue filled circles are for the DM
  run; red filled triangles are for the CSF run; green filled squares are for the AGN
  run. The corresponding open symbols with dotted lines are the results from their
  dark matter components. We further note here that the dark matter component in
  the DM run refer to the first family of dark matter particles (see more details
  in section \ref{4.1}). Solid lines indicate the results from all
  particles inside the cluster.}
\label{fig:rp}
\end{figure*}

We first show the radial profiles of these four parameters: $\eta$, $\zeta$,
$\Delta_r$ and $f_s$, in Fig. \ref{fig:rp} from upper to lower panels respectively.
We select three clusters with different $\eta$ values order as in plot
(>1, $\approx$1 and <1 from the DM run at $R_{vir}$). The results from all
particles are shown with solid lines and filled symbols, while the symbols with
dotted lines from only dark matter particles (DMPs). Different colours
and symbols styles represent different versions of simulations, which are
indicated on the legend of the top middle panel.

It is worth to note at here again that the DM run has two family dark matter
particles: the first (more massive) one shares the same mass and ID to the dark
matter particle in the two hydro-dynamical runs; while the second family only has
its mass the same as the initial gas particles in the two hydro-dynamical runs.
We have verified that this separation in our DM run does not show signs of mass
segregation. This particular set in the DM run allows us to make equal
comparisons to the two hydro-dynamical runs. If it is not particularly noted,
the dark matter particle (DMP) from the DM run refers to the first family (heavier)
particle in the lower part.

To calculate the values of these four parameters at each radius $R_i$, we simply
use the corresponding particles within that radius. However, only particles
inside the spherical shell $R_{i, 0.8} \le R \le R_i$ are used to calculate
the surface pressure energy $E_{s, i}$.
\begin{enumerate}
  \item At inner region, the values of $\eta$ are all larger than 1 for all
  three galaxy clusters, which means that the values of $2T - E_s$ are always larger
  than their potential energy $|W|$. At outer radius (mostly $R \gtrsim 0.6 R_{vir}$),
  $\eta$ becomes more flat for all three clusters and three runs. $\eta$ from
  both the CSF run and the AGN run is normally smaller than from the DM run over
  all radii. However, there is a better agreement between these three
  runs, when only DMP is taken into account, especially at outer regions. It
  means that DMPs are less affected by baryons.
  \item $\zeta$ normally has a value smaller than 1, and shows a declining trend
  from inner to outer radii, which is basically the same saw in $\eta$. In
  agreement with $\eta$, galaxy clusters simulated with baryon models also have
  smaller $\zeta$ values than the DM run. It not surprising that the result from
  DMP is also similar to $\eta$. However, there is a slightly larger disagreement
  between the three runs, especially for the two with $\eta \leq 1$.
  \item The radial profile for the centre-of-mass offset $\Delta_r$
  shows large difference between the three galaxy clusters and the three
  simulation versions. As DMPs contribute the largest mass for galaxy clusters,
  it is not surprising to see that the dotted lines basically follow the solid
  lines. It seems to have less correlation between $\Delta$ and $\eta$, seeing
  from these radial profiles.
  \item It is not surprising that $f_s$ from the CSF run normally has a larger
  value than the other two runs. This is caused by the over-cooling problem, which
  affects not only central galaxies, but also satellite galaxies. Similar to
  $\Delta_r$, there is very little differences between the total (solid lines)
  and DMP (dotted lines) for the $f_s$ profile. $f_s$ for all three clusters
  show a clear increasing trend from inner to outer regions.
  This is simply because the closer to the centre, the higher possibility that
  substructures are destroyed. This trend is anti-correlated with the radial profile
  from $\eta$.
\end{enumerate}

Fro these three example clusters, $\eta$ shows a decreasing trend
from inner to outer radii, which means that galaxy clusters can be highly
un-virialized at their centres than the outer region. In agreement with
\cite{Shaw2006}, $\eta$ at outer radius ($R \gtrsim 0.6 R_{vir}$) becomes more
flat, which means that $\eta$ is primely determined by materials inside
$0.6 R_{vir}$. The in-falling materials at outer region has less effect on $\eta$.
It is interesting to see that baryons give a systematic decreasing
effects on $\eta$ over the whole radii. However, the $\eta$ from DMP seems to be
less affected. Because gravity is the only interaction between dark matter and gas,
and gas only occupy a small mass fraction of clusters with a
smoother distribution, it not surprising to see this results. Because larger
$\sigma$ at fixed radius corresponds to larger $T$, it is also not surprising
to see that $\zeta$ basically follows the trend of $\eta$.

There is no clear trend for the profile of $\Delta_r$. This is because the centre
of mass is largely relying on the mass distributions, especially the substructure
position. However, $f_s$ shows an increasing trend as radius increases.


\subsection{The baryonic effects}

\begin{figure*}
\includegraphics[width=\textwidth]{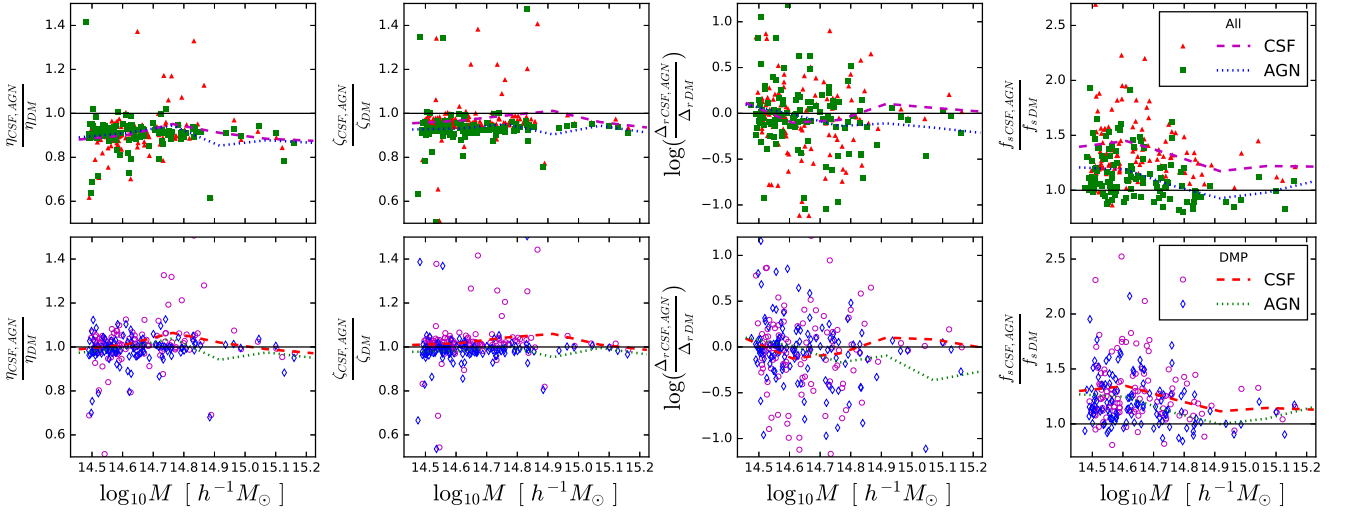}
\caption{From left to right, ratios, as a function of halo masses between the
  CSF/AGN run and the DM run for
  the virial ratio $\eta$, the velocity dispersion deviation $\zeta$, the
  centre-of-mass offset $\Delta_r$ and the substructure mass fraction $f_s$,
  respectively. Lower panels show the same quantities but for DMPs. As indicated
  in the legends on right panels, red triangles with magenta dashed lines (which
  are the mean of data points) are
  coming from the CSF run, while green squares with blue dotted lines
  indicate the AGN run. The reversed color is used for the results from DMPs as
  shown in the legend of the lower right panel. Similar to in Fig. \ref{fig:rp}, DMP
  in DM run refer to the first family of dark matter particles (see more details
  in section \ref{4.1}; while upper panel show the results including all
  particles.}
\label{fig:be}
\end{figure*}

We further investigate baryon effects on the four parameters in Fig. \ref{fig:be}.
To explicitly show and understand this effect, changes of these four parameters
from the DM run to the two hydro-dynamical runs are separated into two rows: the
upper row shows the results from all types of particles; while the lower row is
from DMPs. These results are shown as a function of their halo masses $M_{vir}$.
As shown in the legends of right panels, the different color and style symbols
indicate different simulations; while the different color and style lines are
the mean of data points. The upper row shows the results from all particles;
while the lower row is from DMPs.

Through these comparisons, we find:
\begin{enumerate}
  \item The upper panel from the first column shows the ratio of $\eta$, which
  is calculated with all particles. It is clear that $\eta$ from both the CSF
  run and the AGN run is about 10 per cent lower than the one from the DM run.
  Nevertheless, there is very small difference between the two hydro-dynamical
  runs evident from their mean values. The ratio of $\eta$ shows almost no
  dependence on cluster masses.

  The lower panel shows the results from DMPs. There is almost no difference
  between the two hydro-dynamical runs and the DM run, which is consistent with
  the finding from Fig. \ref{fig:rp}. Although the red dashed line (the CSF run)
  is
  on top of the green dotted line (the AGN run), there is very little difference
  between the CSF run and the AGN run without any dependence on cluster masses.

  \item We show the ratio of $\zeta$ in the second column of Fig. \ref{fig:be}.
  Again, the mean of $\zeta$ from both the CSF run and the AGN run is slightly
  lower ($\sim 0 - 10$ per cent) than from the DM run. However, $\zeta$ from
  the AGN run is closer to the DM results than from the CSF run. Again, this
  ratio shows almost no dependence on cluster masses.

  Similar to the $\eta$ results from DMPs, the mean ratio of $\zeta$ from
  both the CSF run and the AGN run to the DM run is around 1. The difference
  between the CSF run and the AGN run is in consistent the result from the
  upper panel: red dashed line (the CSF run) is always on top of green dotted
  line (the AGN run).

  \item The changes of $\Delta_r$ are shown in log space in the third column.
  Due to its sensitivity to the position of substructures, which seems to be easily
  affected by baryons, there is a large scatter for these data points. However,
  the mean ratio of $\Delta_r$ is around 1 for the CSF run; while this is also
  true for the AGN run at smaller mass, but a slightly smaller $\Delta_r$ than the
  DM run is shown at larger mass.

  Since dark matter normally occupies more than 80 per cent of total cluster mass,
  it is not surprising that the lower panel, which shows the result from DMPs,
  gives very similar results as the upper panel.

  \item We show how $f_s$ changes in the last column. $f_s$ is clear larger
  from the CSF run than the DM run: it increases about 40 per cent at smaller
  cluster mass; it is still about 20 per cent higher at larger cluster mass.
  $f_s$ from the AGN run is 20 per cent lower than from the CSF run;
  20 per cent higher than the DM run at smaller cluster masses, and
  almost no difference between the two runs at $M \gtrsim 10^{14.8} \hMsun$.

  For the changes of $f_s$ from DMPs in the lower panel, we see very similar
  result for the AGN run as in the upper panel. Nevertheless, $f_s$ from the
  CSF run is around 10 per cent closer to the DM run than its result in the upper
  panel over all cluster mass range.

\end{enumerate}

The bottom-left panel in Fig.~\ref{fig:be} shows that $\eta$ is consistent for
DMPs between the three runs, which implies that baryons have little
effect on both the kinetic and potential energy of dark matter if $E_s$ is
ignored. We find a fixed value of $\eta_{DMP}/\eta_{All}$ for
all clusters. This value from the two hydro-dynamical runs is $\sim 10$
per cent higher than from the DM run. This means that baryons have a systematic
change on $\eta$. This is consistent with the finding from the top-left panel of
Fig. \ref{fig:be}. We study this below.


Similar to the findings from radial profiles in Fig. \ref{fig:rp},
$\zeta$ also shows the closest correlation with $\eta$ for the change caused by
baryons. Although the ratio between $\zeta_{AGN}$ and $\zeta_{DM}$ is
very similar to the ratio of $\eta$, $\zeta_{CSF}$ is much closer to $\zeta_{DM}$
than $\eta_{CSF}$ to $\eta_{DM}$. This shows that the over-cooling problem in the
CSF run has more effect on $\zeta$ than $\eta$.


In agreement with Fig. \ref{fig:rp}, baryons have a large influence on $\Delta_r$.
It is not surprising that $\Delta_{r, DMP}$ follows $\Delta_{r, All}$,
and both have a large scatter. However, the mean changes of $\Delta_r$ seem to
rest on 1, except the drop at high mass end from the AGN run. This large scatter
can be caused by the sensitivity of the mass distribution to baryons: 1) galaxy
cluster centers can be changed from the DM run to the two hydro-dynamical runs; 2)
the positions and masses of substructures can be altered by baryons.

$f_s$ from the AGN run seems to suffer a weak baryon effect, except at smaller
mass clusters, which tend to have higher ($\sim 20$ per cent) substructure mass
fraction than the DM run, while the over-cooling problem is more obvious for $f_s$:
substructures from the CSF run are more massive than from the DM run.
\begin{figure*}
\includegraphics[width=\textwidth]{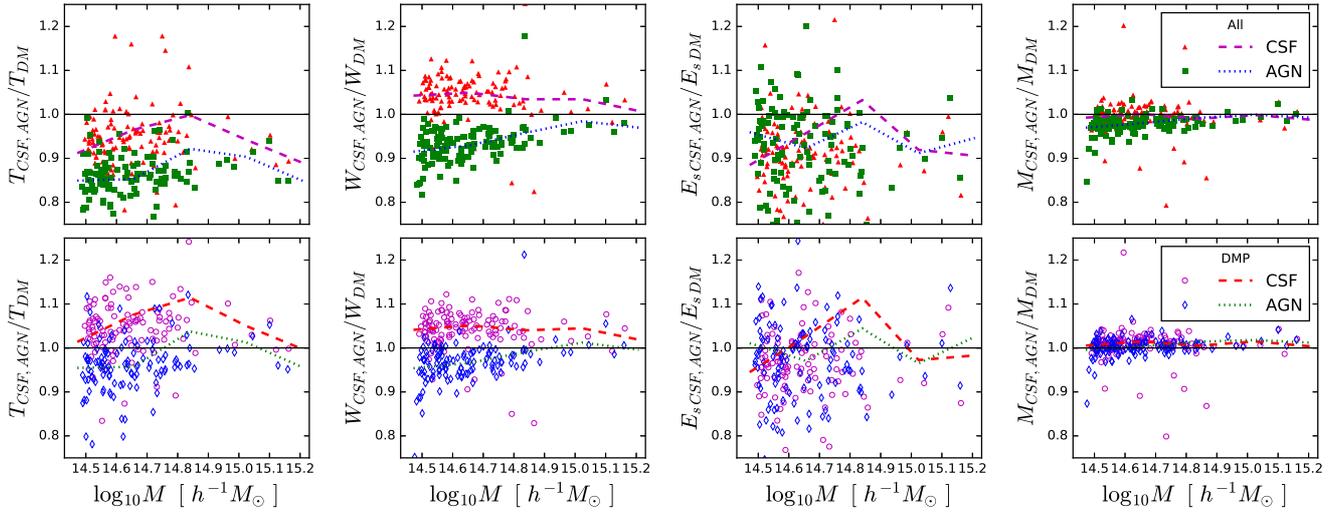}
\caption{ Similar plots as Fig~\ref{fig:be}, but for kinetic energy T, potential
  energy W, surface pressure energy $E_s$ and galaxy cluster mass $M_{vir}$.
  We refer to Fig~\ref{fig:be} and the two legends on the right panels for the
  meanings of the colors, symbols and lines.}
\label{fig:beta}
\end{figure*}

\bigskip

$\eta$ is calculated from kinetic energy T,
potential energy W, and surface pressure energy $E_s$. In \ref{fig:beta}, we study
how $\eta$ is derivative to T, W, $E_s$.
From left to right columns, we show the baryon effects on T, W, $E_s$ and $M_{vir}$
respectively. The upper row shows the results from all particles; while the lower
panel results are coming from DMPs.

The key findings of Fig.~\ref{fig:beta} are summarized as:
\begin{enumerate}
  \item The ratio of kinetic energy T is shown in the first column. Again, the upper
  panel shows the results from all particles. $T_{CSF}/T_{DM}$ is about 0.95.
  However, the mean of this ratio drops to $\sim 0.9$ at both larger and smaller
  mass end; while it reaches $\sim 1.0$ at $M \approx 10^{14.8} \hMsun$.
  $T_{AGN}/T_{DM}$ is about 0.85 - 0.9. For the result coming from DMPs on the
  lower panel, both ratios have a constant shift up of $\sim 10$ per cent.

  \item The second column shows the ratio of potential energy W.
  For the results from all particles in the upper panel, $W_{CSF}/W_{DM}$ is
  about 1.05, which is gradually reaching $\sim 1.0$ at the massive mass end.
  On the contrary, $W_{AGN}/W_{DM}$ is about 0.95, increasing to
  $\sim 1.0$ for the most massive clusters. For the results from DMPs on the lower
  panel, this ratio for the CSF run is almost the same; while the AGN run slightly
  ($\sim 3$ per cent) shift up.

  Although both the CSF and AGN runs tend to have similar virial and dark
  matter masses as the DM run (actually the total mass from the AGN run is a little
  lower than the DM run at smaller halo masses, see the fourth column of this figure
  for more detail), the over-cooling problem in the CSF run tends to result a
  much higher concentration \citep[see more discussion in][]{Cui2016}, and so a
  higher potential energy than the AGN run.

  \item We show ratios of $E_s$ in the third column. We have verified that $E_s$
  only occupies $\sim 20$ per cent of the total kinetic energy T. It means
  that $E_s$ has a minor contribution to $\eta$. As shown on the upper panel,
  the baryon effect on the total $E_s$ is very similar ($\sim 5$ per cent lower
  than the DM run) between the AGN run and the CSF run. It is not surprising
  that DMPs contribute similar to $E_s$ between these three runs, that is shown
  on the lower panel.


\end{enumerate}

From this we conclude that $E_s$ is irresponsible for the baryon effect.
The unchanged $\eta_{DMP}$ for the CSF run is because
baryons have a similar increasing ($\sim 5$ per cent) effect on T and W, while both T
and W seem to be unaffected by baryons for the AGN run.

For the baryon effect on $\eta_{total}$, the key difference is in $T_{total}$.
The drops of $T_{total}$ in
both hydro-dynamical runs are possibly caused by collisional gas, of which
thermal energy is either dissipated due to turbulences and frictions, or locked
up into stars.

\subsection{The classification of relaxed and unrelaxed clusters}

\begin{figure*}
  \includegraphics[width=\textwidth]{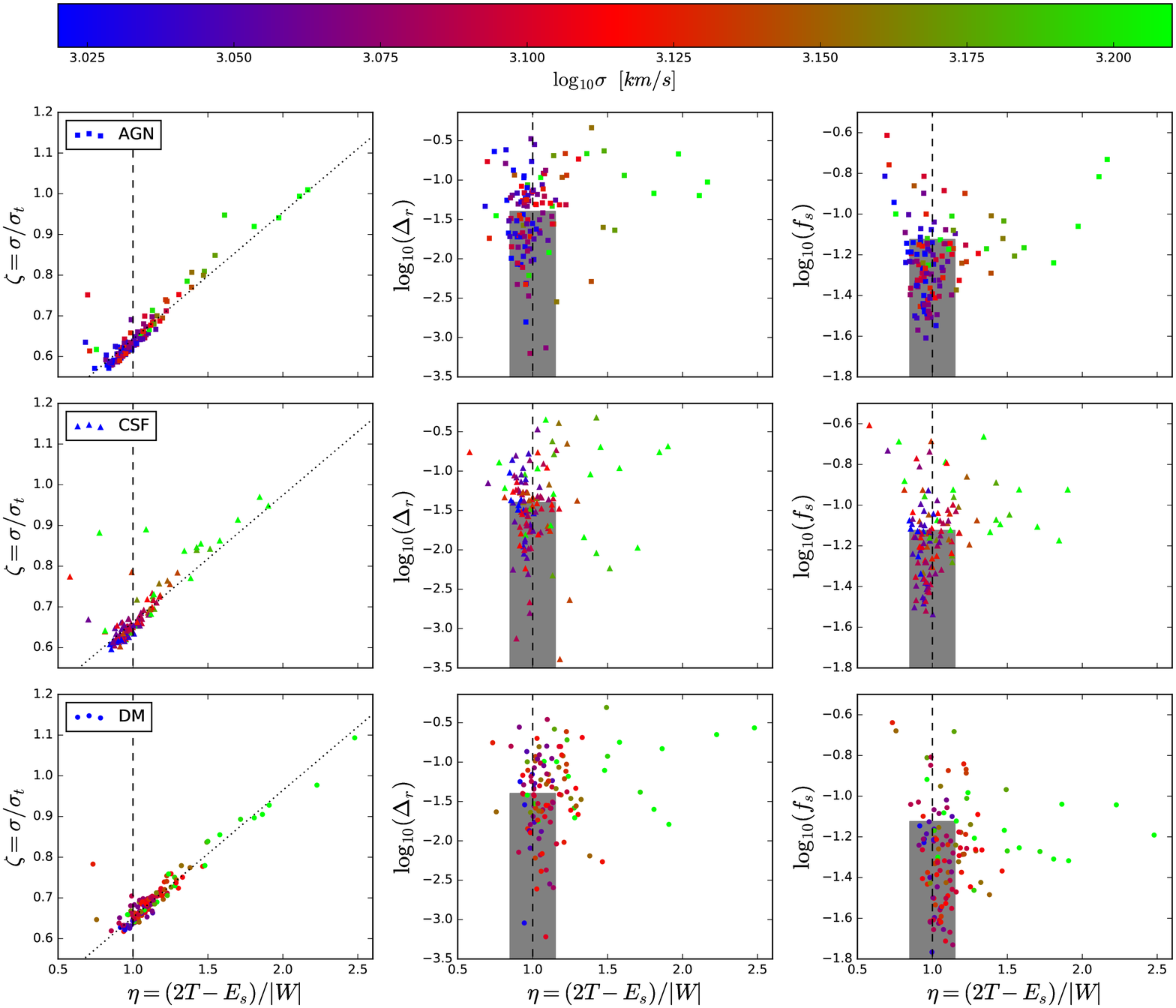}
\caption{Relations between $\zeta$ (left column), $\Delta_r$ (middle column)
  $f_s$ (right column) with the virial ratio $\eta$. From top to bottom panels,
  we show results from the AGN, the CSF and the DM runs. Symbol color is coding
  to its velocity dispersion $\sigma$, which is shown by the colorbar on the
  top of this plot. Dashed vertical lines indicate $\eta = 1$, where the
  cluster is in dynamical equilibrium. Dotted lines on the left column are
  fitting results to data points with a fix slope of 0.312. Galaxy clusters
  located inside grey regions in the middle and right columns can be classified
  as relaxed clusters.}
\label{fig:td}
\end{figure*}

Separating out relaxed clusters from unrelaxed ones is not an easy task.
\citet{Neto2007} adopted $2T/|U| < 1.35$, $\Delta_r < 0.07$ and $f_s < 0.1$ to
select relaxed galaxy clusters. They found $\sim 50$ per cent of halos at
$M_{vir} = 10^{15} \hMsun$ are relaxed. However, they did not take the surface
pressure energy $E_s$ into account in their virial ratio calculation.
\cite{Shaw2006} adopted a slightly narrower limit ($\beta = 0.2$, equivalent to
$|1 - \eta| < 0.2$) to select virial equilibrium halos with $E_s$ in their
$\eta$. With only this criterion, they excluded 3.4 per cent of 2159 halos
($M_{halo} \gtrsim 3 \times 10^{13} \hMsun$) as un-virialized ones.
\cite{Power2012} picked out dynamically
relaxed halos with a slightly smaller $\Delta_r < 0.04$ at $z$ = 0.
From this we conclude that there is no consistency in the literature about
parameter for relaxed halos.


In Fig.~\ref{fig:td}, we investigate relations between these parameters: $\eta$
vs $\zeta$ (left column), $\Delta_r$ (middle column), $f_s$ (right column),
which are normally used for classifying cluster dynamical states.
From top to bottom, we show results from the AGN, the CSF and the DM runs,
respectively. Symbol color encodes the cluster velocity dispersion
$\sigma$, indicated in the top colorbar. Dashed vertical lines show $\eta = 1$,
where clusters are in dynamical equilibrium. Grey regions indicate limits
inside which galaxy clusters are relaxed.

In agreement with Fig.~\ref{fig:be}, there is a good linear correlation between $\eta$
and $\zeta$ shown in the left column of Fig. \ref{fig:td}. This is because
$\sigma$ in $\zeta$ is equivalent to the square root of T in $\eta$, while
$\sigma_t$ is similar to a square root of W. For all three versions of
simulations, $\zeta$ is around 0.65 at $\eta = 1$. After excluding some noisy
data points with $\eta < 0.8$, we find very similar slopes after linear
fitting. Thus, we simply use all data points at the same time to fit, which
results in black dotted lines with a slope of 0.312. This
leads us to propose $\zeta$ as a proxy for $\eta$, which can be deduced
from observation. All particles are used to calculate $\sigma$ and $\zeta$
here. Thus, to apply this relation on observations, one needs to consider the
bias of using galaxies as the velocity dispersion tracer, which has been
investigated and corrected in \cite{Munari2013}, while for cluster mass M in
$\sigma_t$, one can use lensing mass from observation. Using simulations with
mock observation images, \cite{Puchwein2007} have shown that the recovered
lensing mass does not depend on the assumption of hydrostatic equilibrium.
Similar to our proposal, \cite{Puchwein2007} also suggested to use the
difference between dynamically recovered mass from X-ray and lensing mass to
distinguish dynamical states.

Although smaller $\Delta_r$ tends to have $\eta$ closer to 1, there are
clusters that have larger $\Delta_r$ with $\eta \rightarrow 1$. Similarly the same
is true for $f_s$. The virial equilibrium implies that
$<d^2 I / dt^2> = 0$, time-averaged over a period that is long compared to the
local dynamical timescale \citep{Shaw2006, Poole2006}. Therefore, we expect a
roughly symmetric distribution around zero due to those halos that are
oscillating around the virial equilibrium position. These halos with large
$\Delta_r$ and $f_s$ but $\eta \rightarrow 1$ could be still in the process of
settling down to dynamical relaxation, but with a glimpse of equilibrium.

For our limited cluster mass range, we do not see a clear mass dependence on
these parameters in Fig. \ref{fig:td}. However, $\sigma$ shows a weak dependence
on these parameters, especially in the left column, where
higher $\sigma$ value tends to have both higher $\eta$ and $\zeta$ values.
However, this trend is not clear for $\Delta_r$ and $f_s$.

From Fig.~\ref{fig:td}, there is \emph{no bimodal distribution} in any of the
runs for either single or combined parameters.
Data points from all three simulations have a similar
distribution, other than a weak decrease of $\eta$ and a weak increase of $f_s$
from the DM run to the two hydro-dynamical runs.

Applying the selection criteria from \cite{Neto2007}, we find that 70 (78 and
78) out of 123 halos from the DM run (from the CSF run and the AGN run,
respectively) are dynamically relaxed. This gives a similar relaxation fraction
as \cite{Neto2007}. One can visually find out that most of unrelaxed clusters
are cut out by limits from $\Delta_r$ and $f_s$, which is also in agreement
with \cite{Neto2007}. \cite{Power2012} suggested a smaller value of $\Delta_r
\approx 0.04$ to select dynamically relaxed halos \citep[see
also][]{Maccio2007}. Observational results suggest a much lower relaxation
fraction: $\sim 28$ per cent from SDSS survey \citep{Wen2013}; $\sim 16$ per
cent from X-ray selected clusters \citep{Mantz2015}. Thus, we apply restricted
criteria to select out relaxed clusters: $0.85 < \eta < 1.15$
\citep{Knebe2008}; $\Delta < 0.04$ \citep{Power2012}; $f_s < 0.075$. By
applying these thresholds, we select out 41, 43 and 48 dynamical
relaxed clusters from the DM, the CSF and the AGN runs respectively. This gives
a relaxation fraction of $\sim 35$ per cent. 29 ($\sim 65$ per cent) of these
relaxed clusters are cross identified in all three runs; 34 ($\sim 75$ per
cent) of them are cross identified in both the CSF and the AGN runs. In
agreement with the baryon effect on individual parameters, most of halos have
their dynamical relaxation states unchanged. Although AGN feedback
impacts on substructures as well as $f_s$, it plays a minor role in changing
the dynamical state of clusters.

\section{Discussion and Conclusions}
\label{concl}

Using our simulated galaxy cluster catalogue of 123 galaxy clusters from
\citetalias{Cui2015}, we investigated the dynamical state of
clusters in the DM (dark-matter-only) run; the
CSF (gas cooling, star formation and supernova feedback) run and the AGN (with
also AGN feedback) run. These three sets of simulations allow us to explore how
baryons affect cluster dynamical states. We examined four
parameters: the virial ratio $\eta$, the velocity dispersion deviation $\zeta$,
the centre of mass offset $\Delta_r$ and the substructure mass fraction $f_s$,
which are normally used to separate dynamically relaxed clusters from unrelaxed
ones.

The main results are summarised as follows.
\begin{enumerate}
  \item The radial profiles of $\eta$ and $f_s$ become relatively constant at
  outer radius ($R \gtrsim 0.6 R_{vir}$). However, $\Delta_r$ does not show
  such features. It means that we can expect $\eta_{500} \approx \eta{vir}$ and
  $f_{s, 500} \approx f_{s, vir}$. However, this is not applicable for
  $\Delta_r$.

  \item The baryon models (both with and without AGN feedback) have
  a weak effect on $\eta$, which is
  $\sim 10$ per cent lower in the two hydro-dynamical compared to the DM run. This is
  mainly caused by the drop of kinetic energy T with gas dynamics. Therefore,
  $\eta_{DMP}$ shows very similar results between all three runs.

  Baryon models have no impact on $\Delta_R$ for the CSF run; this is also
  true for the AGN run at smaller masses, but there is a slightly smaller $\Delta_r$
  in the AGN run than in the DM run at the higher mass end.

  $f_s$ is about 40 (20) per cent higher in the CSF run than in the DM run at
  smaller (higher) masses, while $f_s$ from the AGN run is 20 per cent
  lower than from the CSF run.

  \item There is good linear correlation between $\eta$ and $\zeta$ for all three
  runs, which encourages us to use $\zeta$ as an indicator of $\eta$, which can
  not be easily measured from observation. Using this relation, one can deduce
  the virial ratio for observed galaxies.

  \item For all the investigated parameters, there is no clear bimodal distribution
  between relaxed and unrelaxed clusters.

\item With more restricted thresholds for $\eta, \Delta_r$ and $f_s$, we find that
  $\sim 35$ per cent of our sample clusters are relaxed, in which $\sim$
  65 per cent are cross identified in all three runs. This means that baryons
  play a minor role in regulating cluster dynamical states.

\end{enumerate}


Using controlled cluster simulations, \cite{Poole2006} quantified the effects of
mergers on the dynamical state of galaxy clusters and showed that dark matter
typically relaxes slightly later than gas. A recent work by \cite{Zhang2016},
who also used controlled cluster simulations but only with adiabatic gas,
investigated baryon effects on merger times. They found merger timescale can be
shortened by a factor of up to 3 for clusters with gas fractions of 0.15,
compared to the one without gas. This indicates that clusters with baryons will
virialize faster than ones without baryons, which is similar to the
finding in \cite{Poole2006}. With galaxy clusters from cosmology simulations,
we only find that baryons decrease the virial ratio by $\sim 10$ per cent from
the DM run, which makes the mean of $\eta$ in the two hydro-dynamical runs much
closer to 1.
Because clusters in
cosmological volume can never be isolated because mergers and in-falling
material are ongoing, their dynamical states can hardly be
exactly in dynamical relaxed. We further note here that the relaxation fraction
seems to be unaffected ($\lesssim 5$ per cent) by baryons. This
could because 1) our cluster sample is not large enough; 2) this relaxation
fraction depends on the arbitrary selection limits.
The total baryon mass fraction is normally around 10 - 15 per cent within galaxy
clusters \citep[e.g.][]{Sun2012, Gonzalez2013, Lagana2013, Borgani2006, Planelles2013}.
It is interesting to see that $\eta$ is dragged down around a similar fraction
by baryons, while its value from dark matter component is almost untouched.
Another unchanged quantity is the linear relation between $\eta$ and $\zeta$,
which urges us to propose a simple fitting function for observers to get
$\eta$ from observed galaxy clusters. However, there is no bimodal distribution
between relaxed and unrelaxed galaxy clusters. It makes a tough task for
choosing the limits for these parameters to select out galaxy
clusters in dynamical equilibrium.

Using different wavelength tracers to determine dynamical states of galaxy
clusters can give different answers. Using photometric data of the Sloan Digital
Sky Survey, \cite{Wen2013} derived the asymmetry, the ridge flatness and the
normalized deviation of a smoothed optical map, which is coming from the
brightness distribution of member galaxies. With their defined relaxation
parameter from the upper three quantities, they found that 28 per cent of 2092
clusters are dynamically relaxed. In X-ray observation, the power ratio and
the centroid shift are normally used to select out dynamically relaxed clusters
\citep[e.g.][]{Boehringer2010,Rasia2013a}. In addition, \cite{Mantz2015}
proposed the symmetry-peakiness-alignment criterion for classifying cluster
dynamical states. With their criterion, they report a relaxation fraction of 16
per cent for their 361 X-ray selected clusters. Combining different wavelength
results could give accurate answers. For example, \cite{Ge2016} has investigated
the dynamical state of two paired clusters under optical, X-ray and radio
emissions; \cite{Rossetti2016} characterized the dynamical states of galaxy
clusters detected with the Sunyaev-Zeldovich (SZ) effect by the Planck and compare
them with their dynamical states derived from X-ray surveys. They found
a slightly higher relaxation fraction from the X-ray sample ($\sim$74 per cent)
than from SZ sample ($\sim$ 52 per cent), which could due to different selection
effects.


The reliability and agreement between these tracers from different
wavelength observations, between different methods, as well as the consistency
with theoretical predictions are still unclear. We will address these questions
with our galaxy cluster sample in the next paper.

\section*{Acknowledgements}

All the figures in this paper are plotted using the python
matplotlib package \citep{Hunter:2007}. This research has made use of NASA's
Astrophysics Data System (ADS), the arXiv preprint server and Wikipedia.
Simulations have been carried
out at the CINECA supercomputing Centre in Bologna, with CPU time
assigned through ISCRA proposals and through an agreement with the
University of Trieste.  WC acknowledges the supports from University of
Western Australia Research Collaboration Awards
PG12105017, PG12105026, from the Survey Simulation Pipeline (SSimPL;
{\texttt{http://www.ssimpl.org/}}) and from iVEC's Magnus supercomputer under
National Computational Merit Allocation Scheme (NCMAS) project gc6.
WC, CP, AK, GFL, and GP acknowledge support of ARC DP130100117.
CP, AK, and GFL acknowledge support of ARC DP140100198.
CP acknowledges support of ARC FT130100041.
SB and GM acknowledge support from the PRIN-INAF12 grant 'The Universe in a
Box: Multi-scale Simulations of Cosmic Structures', the PRINMIUR 01278X4FL
grant 'Evolution of Cosmic Baryons', the INDARK INFN grant and 'Consorzio per la
Fisica di Trieste'. AK is supported by the {\it Ministerio de Econom\'ia y
  Competitividad} (MINECO) in Spain through grant AYA2012-31101 as well as the
Consolider-Ingenio 2010 Programme of the {\it Spanish Ministerio de Ciencia e
  Innovaci\'on} (MICINN) under grant MultiDark CSD2009-00064.  He further thanks
Luna for lunapark. GP acknowledges support from the ARC Laureate program of
Stuart Wyithe.

\bibliographystyle{mnras}
\bibliography{bibliography}

%

\bsp
\label{lastpage}
\end{document}